\documentclass[twocolumn]{article}
\usepackage{graphicx}
\usepackage{eqnarray}
\usepackage{amssymb}
\usepackage{anysize}
\usepackage{amsfonts,amssymb,latexsym,amsmath,amscd}
\marginsize{0.7in}{0.7in}{0.7in}{0.7in} \textheight=9.7in

\begin{document}
    \title{A Nano-Quantum Photonic Model for Justification of Dispersion in Single Crystal
    Film of NPP}
    \author{Hassan Kaatuzian,  AliAkbar Wahedy Zarch \footnote{Material presented in this
    paper is a part of Ali Akbar Wahedy Zarch's work on his
    thesis towards Ph.D degree. Dr. Hassan Kaatuzian is his advisor on
    thesis.}\\
    Photonics Research Laboratory\\Electrical Engineering Department\\AmirKabir University
     of Technology, Tehran, IRAN\\
     hsnkato@aut.ac.ir\\fatemahali@noavar.com}

    \maketitle

    \begin{abstract}
       In this paper, we present a nano-quantum photonic model for justification of normal
       dispersion in a thin crystal film of NPP. In this method, we assume a laser beam
       consists of a flow of energetic particles. By precise analyzing of photon interaction
       with $\pi$-electron system of benzene ring in NPP crystal, we will attain refractive
       index (RI) in any wavelength and compare the results with experimental data.
    \end{abstract}

    \vspace{2pc} \noindent{\it Keywords}: $\pi$-electron system, photon-electron interaction,
    dispersion, NPP.\\
    \section{Introduction}
    Organic optical materials like MNA, NPP, MAP have a high figure of merit in optical
    properties in comparison with inorganic optical materials such as LiNbO$_{3}$, ADP,
    KDP and GaP \cite{1}. Additionally N-(4-nitrophenyl)-L-prolinol (NPP) after MNA has the highest
    figure of merit between organic nonlinear optical materials \cite{2}. Thus it is used for electro-optic
    and nonlinear optic applications \cite{1,2,3,4,5,6,7,8,9,10,11}. It may say that
    refractive index (RI) is the most effective element in optical phenomena.
    Ledoux et al. first proposed a Sellmeier set for RI of NPP, based on measured
    refractive indices\cite{3}. Banfi, Datta and co-workers design a more accurate
    Sellmeier set for RI of NPP for nonlinear optical studies \cite{4,5,6}; these data based on
    classic and macroscopic measurements and approaches. Some authors
    explain RI in molecular bases \cite{12,13,14}, although no data on real
    material exist in this literature. Our purpose in this paper is
    the explanation of RI or dispersion effect of a real material,
    a single crystal film of NPP, in an approach that name
    nano-quantum photonic \cite{15,16,17}. This approach is based on four elements: 1-
    quasi-quantum principle for justification of phenomenon in
    molecular scales, 2- Knowledge of crystal network and its space shape, 3-
    Short range intramolecular and intermolecular forces, 4-
    Monte-Carlo time domain simulation. We suppose a laser beam is a
    flow of photons when passes through single crystal film, interacts
    with delocalization $\pi$-electron system of NPP molecule and delays
    the photon in every layer. By precise calculation of these
    retardation in every layer, we obtain refractive index (RI) in
    specific wavelength. The results obtained from this method are
    agreeable to experimental data. Our previous work was about MNA
    aromatic crystal, with a simpler structure than NPP molecule,
    \cite{17,18}. By simulation on more complex crystal, NPP, and
    considering the agreement of simulation results with
    experimental and Sellmeier data, the validity of assumptions and simulations will be
    proved. Hence, first we introduce crystal and molecular structure of NPP
    and comparison to Benzene molecule, after which we explain normal
    dispersion formula according to classical physics,
    electromagnetism, quantum electrodynamics and our microscopic
    model. In fourth section, we simulate and calculate the
    probability density function (PDF) of electron presence in benzene
    ring of NPP molecule, approximately. In fifth section, we express
    detailed calculation of RI. In sixth part, we clarify how to
    obtain RI for NPP crystal and compare simulation results with experimental and Sellmeier
    data.

    \section{Crystal and Molecular structure of NPP}

        Organic molecular units and conjugated polymer chains possessing  $\pi$-electron
        systems usually form as centrosymmetric structures  and thus, in the electric
        dipole approximation, would not show any nonlinear second order optical properties.
        The necessary acentric may be provided by first distorting the  $\pi$-electron system
        by interaction with strong electron donor and acceptor groups \cite{19}.
        In NPP molecule nitro group acts as an acceptor and prolinol group on the other side of benzene
        ring acts as a week donor (see fig. 1).
        \begin{figure}
            \centering
            \includegraphics[scale=0.4]{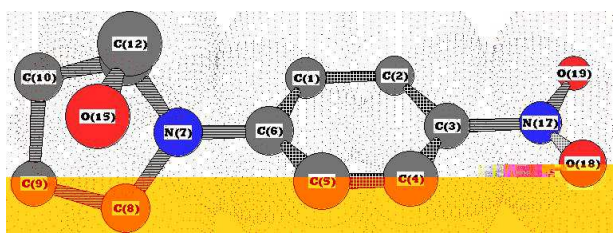}
            \caption{The molecular compound of NPP.}\label{Fig1}
        \end{figure}
        NPP $(C_{11}H_{14}N_{2}O_{3})$ (Fig.1) crystallizes in the solid
        state in an acentric monoclinic (with space group $P2_{1}$) structure and
        their parameters are:a=5.261$A^{\circ}$,b=14.908$A^{\circ}$,c=7.185$A^{\circ}$
        ,$\beta$=105.18$^{\circ}$.
        The volume of unit cell is V=543.8($A^{\circ^{3}}$).
        The molecular weight of NPP is M=222 and the crystal with two molecules in the unit
        cell has a calculated density D=1.36 gm.cm$^{-1}$ \cite{1}. The melting point of NPP
        is 116$^{\circ C}$ and in the wavelength range of 0.48 to 2$\mu$ is transparent
        \cite{2}. The most
        interesting property of NPP crystal
        is the proximity of the mean plane of molecule with the crystallographic
        plane (101); the angle between both of these planes being 11$^{\circ}$.
        Nitro group of one molecule in downward connects to prolinol group in upper
        by hydrogen bonding. The angle between b orientation of crystal and
        N(1)-N(2) axis is equal to 58.6$^{\circ}$ \cite{1}. Fig. 2 shows
        \begin{figure}
            \centering
            \includegraphics[scale=0.4]{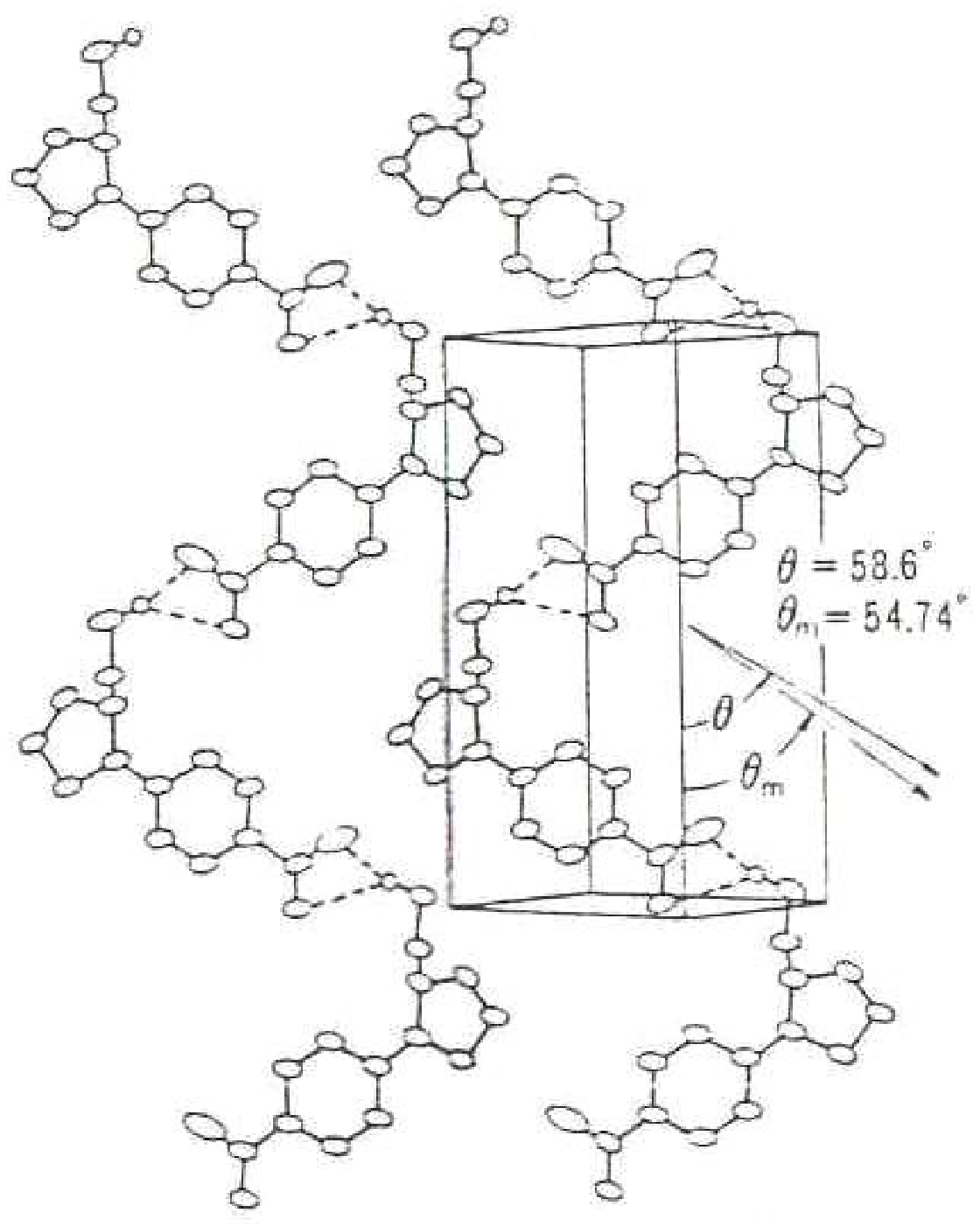}
            \caption{The crystal packing of NPP \cite{2}.}\label{Fig2}
        \end{figure}
        the crystal packing
        of the NPP. For accurate and valid simulation, these properties and angles have
        to be exerted. For benzene molecule, benzene ring is a circle (see fig. 3);
        \begin{figure*}
            \centering
            \includegraphics[scale=0.7]{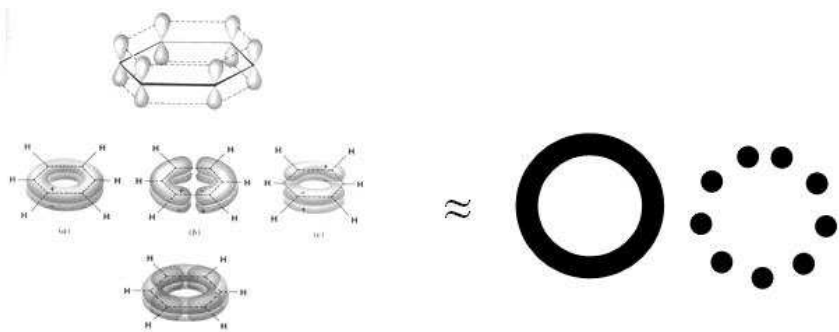}
            \caption{Electron cloud for Benzene molecule that obtained from Huckle theory
            \cite{20} and its approximation.}\label{Fig3}
        \end{figure*}
        Fig. 4 demonstrates
        \begin{figure}
            \centering
            \includegraphics[scale=0.35]{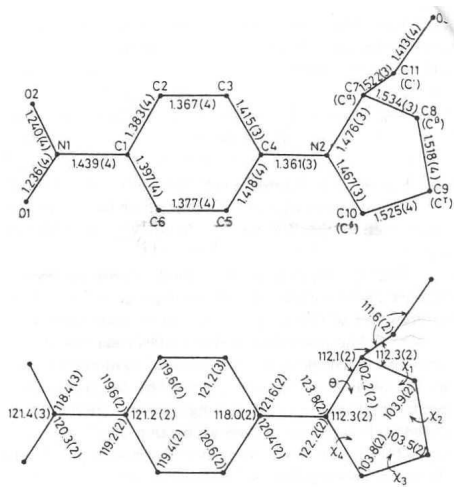}
            \caption{The bond lengths and angles of NPP molecule \cite{1}.}\label{Fig4}
        \end{figure}
        the bond lengths of the NPP molecule. As we see in this figure the bond lengths
        in benzene ring are not same. In our simulation, for similarity
        we consider an ellipse correspond to circle for electron cloud.
        We obtained $\varepsilon$=0.26 for that ellipse
        from simulation, fig. 5 shows this comparison.
        \begin{figure}
            \centering
            \includegraphics[height=3in,width=3in,scale=0.38]{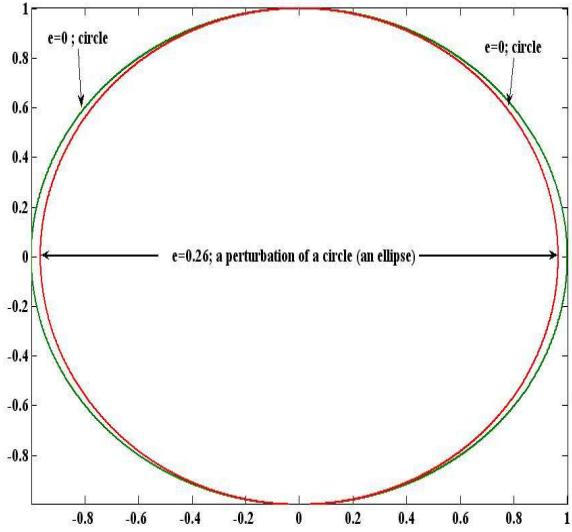}
            \caption{The comparison between circles for Benzene
            molecule electron cloud and NPP molecule electron cloud, (approximately).}\label{Fig5}
        \end{figure}
    \section{Classic and microscopic model for dispersion}
        Cauchy suggested an equation that performs a relation between
        wavelength and RI in form \cite{21}:
        \begin{equation}
            n-1=\frac{A}{\lambda^{2}}+\frac{B}{\lambda^{4}}+\frac{C}{\lambda^{6}}+...
        \end{equation}
        If we have three RI in three wavelength according to experimental data, then
        we have RI in any wavelength in transparent range of that
        substance. Ledoux et.al \cite{3}proposed a Sellmeier formula
        for NPP crystal in the form:
        \begin{displaymath}
            n^{2}=A+\frac{B}{1-\frac{C}{\lambda^{2}}}+D\cdot\lambda^{2}
        \end{displaymath}
        and Datta et al. \cite{5} demonstrated a more accurate of it in the form:
        \begin{displaymath}
            n^{2}=A+\frac{B}{1-\frac{C}{\lambda^{2}}}+\frac{D}{1-\frac{E}{\lambda^{2}}}
        \end{displaymath}
        that A,B,C,D and E is exerted from table 1.
        \begin{table*}
             \centering
             \caption{Sellmeier data of NPP}\label{Table1}
             \begin{tabular}{|ccccc|ccccc|}
             \hline
             \multicolumn{5}{|c}{Ledoux et al.\cite{3} selmeier form}& \multicolumn{5}{c|}{Datta and Banfi et al.\cite{5}
             selmeier form}\\ \hline\hline
             $n$     & $A$ & $B$ & $C$ & $D$ & $A$ & $B$ & $C$ & $D$ & $E$ \\ \hline
             $n_{x}$ & 2.3532 & 1.1299 & 0.1678 & 0.0392 & 2.41704 & 1.08674 & 0.16933 & -0.34200 & 10\\ \hline
             $n_{y}$ & 2.8137 & 0.3655 & 0.2030 & -0.0816 & 2.76667 & 0.37156 & 0.20289 & 0.47880 & 10\\ \hline
             $n_{z}$ & 2.1268 & 0.0527 & 0.1550 & -0.0608 & 2.19965 & 0.00457 & 0.17160 & 0.59363 & 10\\ \hline
             \end{tabular}
        \end{table*}
        These formulas that derived from Sellmier equation, gives no microscopic perspective of
        interactions of photon-electron when light passes through the material;
        additionally in sub-micron scales, RI loses its stabilization and these
        equations are not useful \cite{15}.\\ \indent
        Electromagnetism have also a dispersion equation:
        \begin{equation}
            n-1=\frac{q_{e}^2}{2\varepsilon_{\scriptscriptstyle{0}}m}\sum_{k}\frac{N_{k}}{\omega_{k}^{2}-
            \omega^{2}+i\gamma_{\scriptscriptstyle{k}}\omega}
        \end{equation}
        where there are $N_{k}$ electrons per unit of volume, whose natural frequency
        is $\omega_{k}$ and whose damping factor is $\gamma_{k}$,
        \cite{8}.\\ \indent
        Quantum Electrodynamics attains a famous dispersion relation in form of:
        \begin{equation}
             n^{2}-1=\frac{N}{\hbar\varepsilon_{\scriptscriptstyle{0}}}
             \sum_{\nu}\{\frac{(e_{\scriptscriptstyle{1}}\cdot\mu_{\scriptscriptstyle{l\nu}})
            (e_{1}\cdot\mu_{\scriptscriptstyle{\nu l}})}{\omega_{\scriptscriptstyle{\nu l}}+\omega}+
            \frac{(e_{\scriptscriptstyle{1}}\cdot\mu_{\scriptscriptstyle{l\nu}})
            (e_{\scriptscriptstyle{1}}\cdot\mu_{\scriptscriptstyle{\nu l}})}{\omega_{\scriptscriptstyle{\nu l}}-\omega}\}
        \end{equation}
        by Density Matrix Method where:
        \begin{displaymath}
            W_{n}-W_{m}=\omega_{nm}
        \end{displaymath}
        where W is the energy level and
        \begin{displaymath}
            \mu_{\scriptscriptstyle{ik}}=i\frac{e}{m\omega_{\scriptscriptstyle{ik}}}p_{\scriptscriptstyle{ik}}
        \end{displaymath}
        that introduce the notation:
        \begin{displaymath}
            e_{i}=(e_{x}^{(i)},e_{y}^{(i)},e_{z}^{(i)});\ \ \ \ \ \ i=1,2
        \end{displaymath}
        for unit of the polarization of the incident and scattered light, and
        \begin{displaymath}
            r=(x,y,z)
        \end{displaymath}
        for position vector of the electron, and
        \begin{displaymath}
            p=(p_{x},p_{y},p_{z})
        \end{displaymath}
        for the electron momentum vector \cite{22}. This formula is the simplest
        form in thermal equilibrium and is very complex in real world.
        This means that for attaining RI from quantum electrodynamics theory,
        we require to measure electron momentum vector and certainly, these
        elements can be calculated on the basis of semiclassical
        arguments.\\ \indent
        Now, we suggest a microscopic model for RI and wavelength relationship.
        In this pattern, average photon flux is related to classic electromagnetic
        concepts like: intensity, power and energy. For example if we have a monochromatic
        laser beam with frequency $\nu$ and intensity I, then we may attain average photon
        flux from relation \cite{23}:
        \begin{equation}
            \label{eq4}
            \phi=\frac{I}{h\nu}
        \end{equation}
        now if we assume thin single crystal film of NPP in $"b"$ direction of
        crystal (or $"z"$ axis) radiated by a He-Ne laser with: $\lambda$=633 nm,
        average power=10mw and beamwidth=20 microns, then from (\ref{eq4}) average photon
        flux is equal to $10^{22} photons/(s-cm^{2})$ that signifies in every second
        $10^{22}$ photons arrive to each centimeter square. Moreover from data of
        crystal in section $(II)$ in every 36.5$(A^{\scriptstyle{\circ^{2}}})$ in z direction, one NPP molecule exists.
        Therefore in every second $36.5\times10^{6}$ photons interact with any molecule or
        in other words in every 27ns, one photon interacts with any NPP molecules.
        In each interaction between photon and electron in every layer of crystal,
        we suppose delay time equal to $\tau_{i}$ (ith layer of crystal). Total retardation
        time for m layers in crystal region is equal to:
        \begin{displaymath}
            \sum_{i=1}^{m}\tau_{i}
        \end{displaymath}
        Consequently required time
        for photon transmission in L length of crystal is equal to $\tau$, achieved from
        relation:
        \begin{equation}
            \tau=\frac{L}{\frac{c_{0}}{n}}=\frac{nL}{c_{0}}=\frac{L}{c_{0}}+\sum_{i=1}^{m}\tau_{i}\\
        \end{equation}
        or
        \begin{equation}
            n-1=\frac{c_{0}}{L}\sum_{i=1}^{m}\tau_{i}
        \end{equation}
        where $c_{\scriptscriptstyle{0}}$ is velocity of light in vacuum. By using this relation, we can relate
        macroscopic quantity $n$ to microscopic quantity $\tau_{i}$. According to (3) if, $n=3, L=3\mu$, consequently
        $\Sigma\tau$ is equal to $10^{-14}$sec. Because in b direction of crystal in $3\mu$ length, approximately 4024
        molecules exist, therefore the average quantity of $\tau$:
        \begin{displaymath}
              \overline{\tau}=\frac{\sum\tau}{N}
        \end{displaymath}
        is in order of $10^{-18}$ sec. From this quantity and previous quantity that in every 27ns,
        one photon interacts with one NPP molecule, we conclude that in every moment, just one photon
        interacts with one molecule. Because NPP molecule has delocalization electrons,(or $\pi$-electron system),
        in benzene ring, that photon interacts with this electron type and it is annihilated \cite{24}. We call this photon,
        a successful photon, (that does not produce phonon). By this method, we are able to simulate microscopic
        perturbation of RI for NPP successfully. Equations (1), (2) and (3) use average quantities in frequency domain,
        but in (6), we utilize momentary quantities in time domain that perturb on average values.
        In our nano-quantum photonic model, we may estimate photon-electron mutual actions in very short time,
        on the order of "$attosecond", (10^{-18}s)$. We claim that our approach is applicable for
        optical material that their crystal structure is known, especially aromatic optical crystal.
        Because, they have a delocalization electron and successful photon interact with it. For inorganic optical
        material if we know active electron in optical phenomenon, we are able to use this method. In our approach,
        we do not have any measurement and we have simulation and estimation with an ideal observer.
        This is the first time that we have focused on a real material for approximating its constitutive
        parameters (like $n,\varepsilon,\sigma, \mu$). Neither in classical nor in Q.M. approach we only
        have typical material. In Q.E.D, we have not seen yet practical or predictable data for a real material.
        Our approach is quasi-classic and is similar to the Compton effect but in greater wavelengths. We have used
        some approximations for problem solution.

        \section{PDF calculation of $\pi$-electron in benzene ring of NPP}
        To obtain $\pi$-electron wavefunction for benzene molecule the Schr\"{o}dinger equation may be solved.
        Since this is very complex, this cannot be done exactly, an approximated procedure known as H\"{u}ckle method
        must be employed. In this method, by using H\"{u}ckle Molecular-Orbital (HMO) calculation, a wave function
        is formulated that is a linear combination of the atomic orbitals that have overlapped \cite{20} (see Fig.3)
        (this method is often called the linear combination of atomic orbitals); that is:
        \begin{displaymath}
            \label{eq9}
            \Psi=\sum_{i}C_{i}\Phi_{i}
        \end{displaymath}
        where the $\Phi_{i}$ refers to atomic orbitals of carbon atoms in the ring and the summation is over the six C
        atoms. The Schr\"{o}dinger equation for a delocalized electron is:
        \begin{displaymath}
            [\frac{-\hbar^{2}}{2m}\nabla^{2}+\sum_{i}V_{i}]\Psi=E\Psi
        \end{displaymath}
        that $V_{i}$ is the atomic potential of $ith$ atom. Following the common approach in quantum mechanics,
        we multiply the above equation from the left by $\Phi_{i}$ and integrate over space.
        By solving this set of algebraic equations, the $C_{i}$ coefficient, the $\Phi_{i}$ wavefunction and
        energy E will be obtained, (Fig. 3). Thus, the $|C_{i}|^{2}$ is the probability of the $\pi$-electron at $ith$
        atom. Thus:
        \begin{displaymath}
        \label{eq11}
            |C_{1}|^{2}+|C_{2}|^{2}+|C_{3}|^{2}+|C_{4}|^{2}+|C_{5}|^{2}+|C_{6}|^{2}=1
        \end{displaymath}
        In the case of Benzene molecule:
        \begin{displaymath}
          |C_{i}|^{2} =\frac{1}{6}
        \end{displaymath}
        as followed from the symmetry of the ring \cite{20,25,26}.
        But for NPP molecule there isn't such as Benzene molecule. NPP is a polar molecule. Nitro $(NO_{2})$ is more
        powerful electronegative compound than prolinol and pulls $\pi$-electron system; consequently, the probability
        of finding $\pi$-electron system at various carbon atoms of main ring isn't same and the probability of finding
        $\pi$-electrons near the Nitro group is greater than near the prolinol group. Therefore there is no symmetry
        for NPP and electron cloud is spindly or oblong, (similar to dom-bell) (Fig. 6).
        \begin{figure}
            \centering
            \includegraphics[scale=0.7]{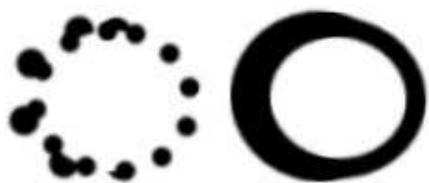}
            \caption{Assumed $\pi$-electron orbit of NPP molecule that obtained from fig. 5 for Benzene molecule
            (approximately).}\label{Fig6}
        \end{figure}
        We estimate this form of electron cloud by an ellipse
        that our calculations would be uncomplicated. According to Bohr atomic model, we assume $\pi$-electron is
        rotating around the center of positive charge in agreement with Kepler law. The positive charge
        in this approximation is virtual.\\ \indent
        For attaining probability of electron presence on an orbit (fig. 7),
        \begin{figure}
            \centering
            \includegraphics[scale=0.7]{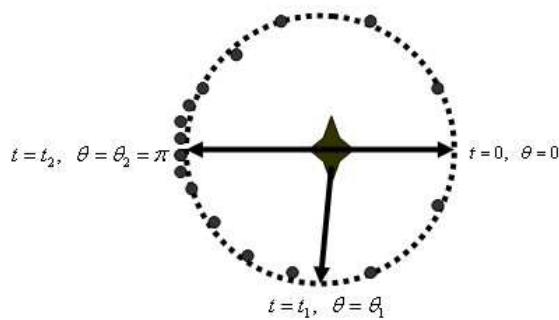}
            \caption{Bohr atomic model and Kepler law for $\pi$-electron system on an orbit.}\label{Fig7}
        \end{figure}
        we say, T time is required by radial vector to sweep total $\pi.u.v$
        interior area of ellipse (u and v are semimajor and semiminor axis of ellipse respectively),
        in t times, this radial vector sweeps
        \begin{displaymath}
          \pi.u.v.\frac{t}{T}
        \end{displaymath}
        area of ellipse, (see fig. 7). If t is the time, that electron sweeps $\theta$ radian of orbit
        then t is obtained from this relation \cite{27}:
        \begin{equation}
            \label{eq12}
            t=\frac{T}{2\pi}\{2\arctan(\sqrt{\frac{1-\varepsilon}{1+\varepsilon}}\tan(\frac{\theta}{2}))-
            \frac{\varepsilon.\sqrt{1-\varepsilon^{2}}.\sin(\theta)}{1+\varepsilon.\cos(\theta)}\}
        \end{equation}
        Where $\varepsilon$ is ellipse eccentricity. By using this relation, we attain the required
        time for electron to traverse from $\theta$ to $\theta+d\theta$ (dt) and it is divided
        by total time T. By this approach, we can determine the PDF approximately. Fig. 8 and Fig. 9 show
        \begin{figure}
            \flushleft
            \includegraphics[scale=0.37]{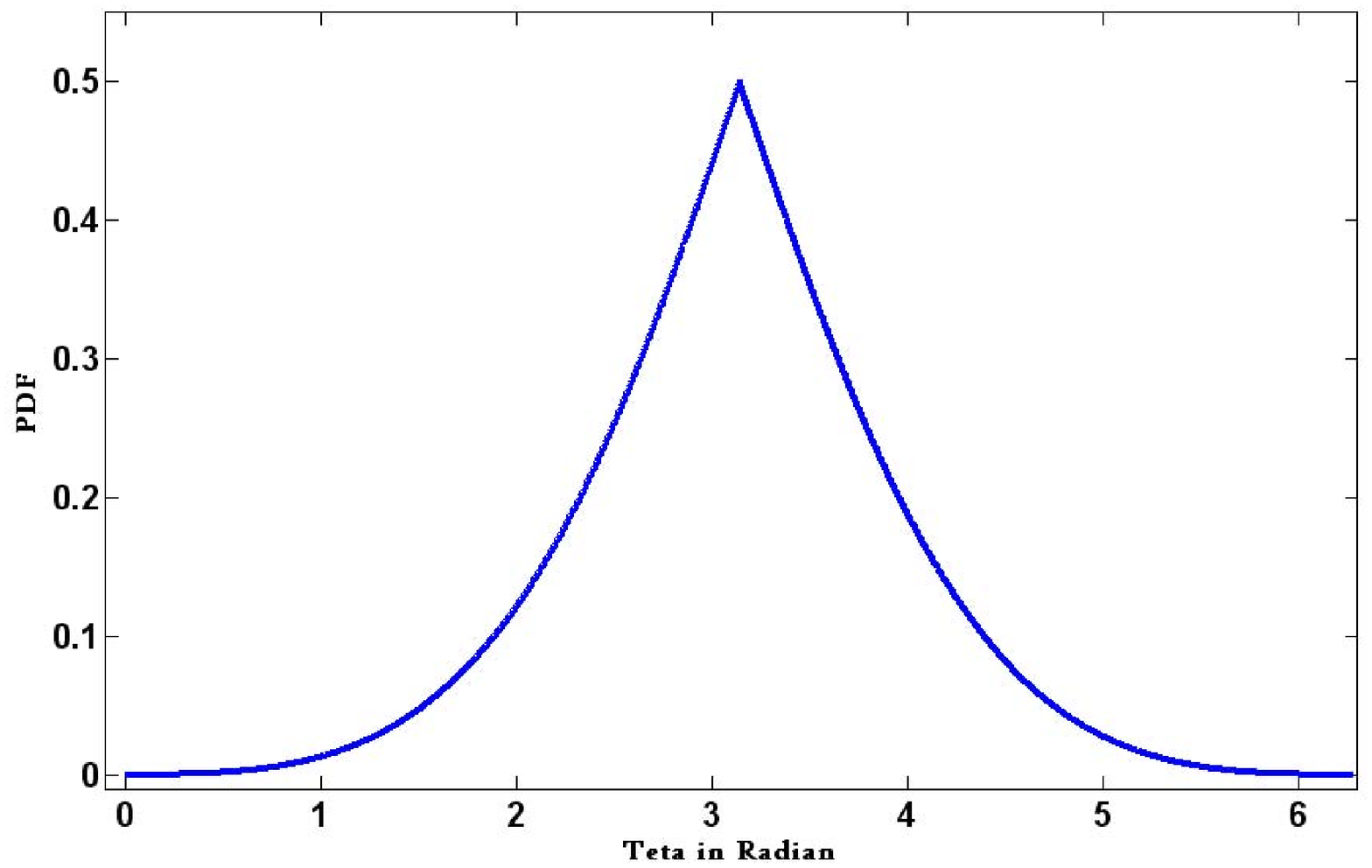}
            \caption{PDF of presence probability of $\pi$-electron around the
            assumed orbit in fig. 7;}\label{Fig8}
        \end{figure}
        \begin{figure}
            \flushleft
            \includegraphics[scale=0.37]{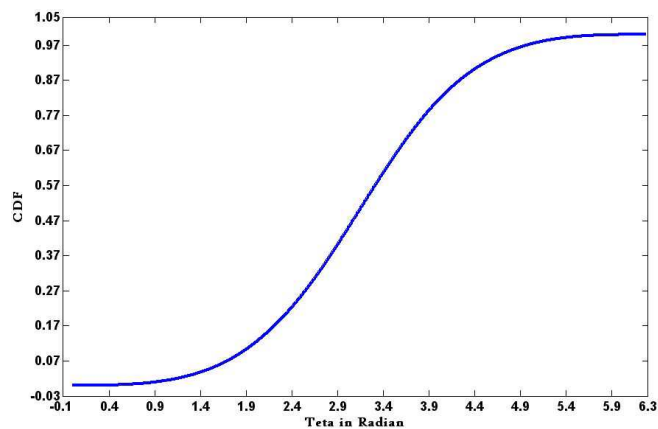}
            \caption{CDF of presence probability of $\pi$-electron around the
            assumed orbit in fig. 7;}\label{Fig9}
        \end{figure}
        PDF and CDF(Cumulative Density function) respectively for presence probability of $\pi$-electron
        around the assumed orbit in fig. 7. Such as shown in this figure the PDF in apogee
        (near the Nitro group), is maximum and in perigee (near the prolinol group) is minimum.

        \section{$\tau$ Calculation}

        The angle between Y vector and charge transfer action (N$_{1}$-N$_{2}$) is 58.6$^\circ$ and X and Z
        axis is perpendicular to Y (Fig.10). We consider propagation along Z direction. We spot a photon interacts
        \begin{figure}
            \centering
            \includegraphics[scale=0.5]{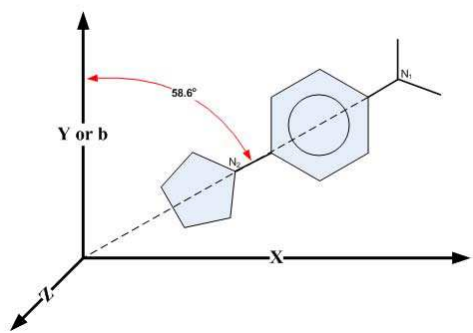}
            \caption{X, Y and Z axis and NPP molecule in dielectric frame.}\label{Fig10}
        \end{figure}
        with $\pi$-electron of NPP in first layer. After interaction, this photon gives
        its energy to electron and is annihilated. Electron absorbs energy and digresses in direction of photon momentum.
        Electron with photon energy, may not be unbounded and after arriving to apogee of digression, it returns back to
        main orbit, because the photon energy is equal to $h\nu=1.96ev$ (h is Planck's constant and $\nu$ is frequency of
        laser beam) whereas energy for ionization is greater than $5ev$. When electron returns to main elliptical orbit
        one photon is produced. The time coming up and down is $\tau$ delay time. This photon after freedom goes to second
        layer in direction of annihilated photon (nonce, we assume the polarization doesn't change), in second layer
        this photon interacts with another delocalization $\pi$-electron
        certainly, because the effective range of photon is approximately equal to its wavelength and is very greater than
        the distance between molecules. This action is repeated for each layer. The location of photon-electron
        interaction is significant in every molecule and it is effective on $\tau$ quantity directly.
        We assume that interacting photon has a field in X direction and electron subject to virtual positive charge
        center. According to fig.11 we have:
        \begin{figure}
            \centering
            \includegraphics[scale=0.7]{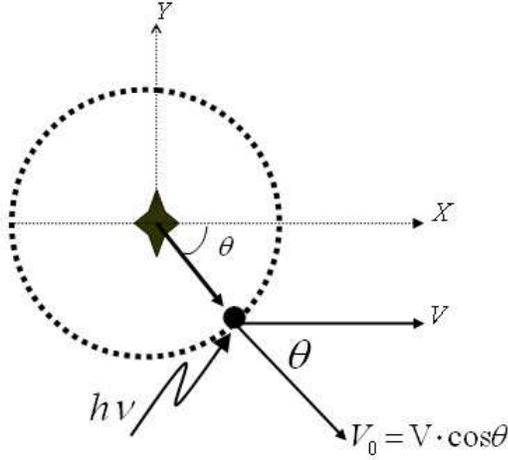}
            \caption{Calculation of $\tau$}\label{Fig11}
        \end{figure}
        \begin{displaymath}
            h\nu=\frac{1}{2}mV^{2}=\frac{1}{2}m(\frac{V_{ox}}{cos(\theta)})^{2}
            \end{displaymath}
        and consequently:
        \begin{displaymath}
            V_{ox}=\sqrt{\frac{2h\nu}{m}}.cos(\theta)
        \end{displaymath}
        that $\frac{1}{2}mV^{2}$ is the kinetic energy of electron after interaction with photon, and columbic force, F,
        is equal to:
        \begin{displaymath}
           F=K\frac{Ze^{2}}{r^{2}}=ma
        \end{displaymath}
        where $e, a$   and $m$ are charge, acceleration and mass of electron respectively. $Ze$ is positive charge
        equivalent in effective center of positive charge. Consequently the time required for electron to go up and
        down is equal to:
        \begin{equation}
            \label{eq16}
            \tau_{x}=\frac{V_{ox}}{a}=\frac{\sqrt{2h\nu.m}.cos(\theta)}{KZe^{2}}.r^{2}
        \end{equation}
        and similarly:
        \begin{equation}
            \label{eq16}
            \tau_{y}=\frac{V_{ox}}{a}=\frac{\sqrt{2h\nu.m}.sin(\theta)}{KZe^{2}}.r^{2}
        \end{equation}
        that
        \begin{displaymath}
            r=\frac{(1-\epsilon^{2}).u}{1+\epsilon.cos(\theta)}
        \end{displaymath}
        where $\epsilon$ is the elliptical eccentricity
        and u is the semimajor axis of the ellipse. By performing $\tau_{x}$ and $\tau_{y}$ separately in every layer
        and using (6), we attain $n_{x}$ and $n_{y}$ separately.

        \section{Attaining of $n_{x}$ and $n_{y}$ for NPP crystal}
\begin{figure}
            \centering
            \includegraphics[scale=0.37]{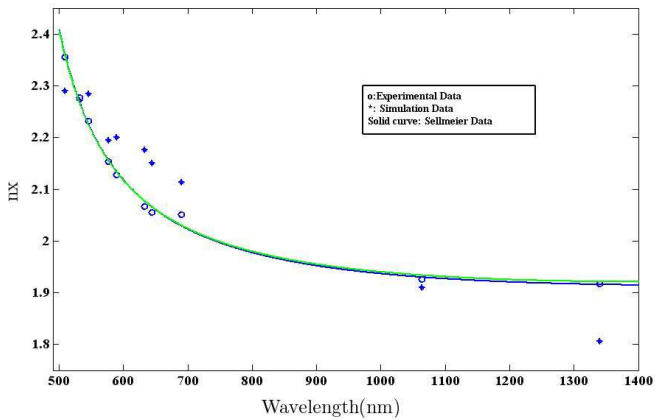}
            \caption{Comparison of experimental data, Simulation data and Sellmeier data for $n_{x}$}\label{Fig12}
        \end{figure}

\begin{figure}
            \centering
            \includegraphics[scale=0.37]{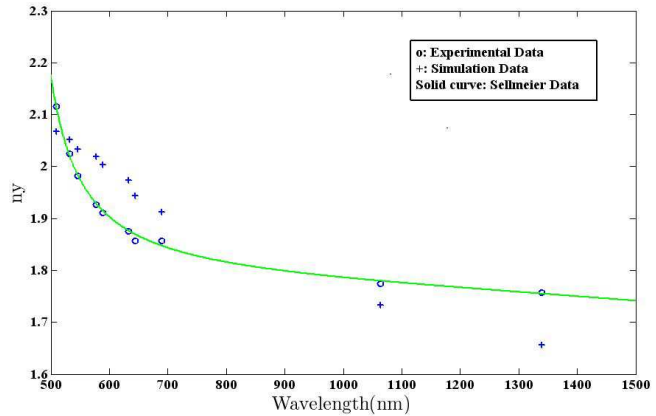}
            \caption{Comparison of experimental data, Simulation data and Sellmeier data for $n_{y}$}\label{Fig13}
        \end{figure}

        We are now in a position to take up $n_{x}$ and $n_{y}$ by Monte-Carlo method, then we simulate random
        number producer using $MATLAB$ program. This program produces PDF quantities explained in section (4)
        and relates each of them to every molecule. These values are indexing $\pi$-electron positions in each
        layer, by assumption a reference point (see fig. 7). Additionally we have used a $MATLAB$ program for Monte-Carlo
        simulation. The inputs of this program are:\\ \indent
        1.  The wavelength of incident optical beam in which we want to obtain refractive index of NPP
        crystal;\\ \indent
        2.  $h, m, q, k=\frac{1}{4\pi.\varepsilon}, c_{\scriptscriptstyle{0}}$ that are Planck's constant,
        electron rest mass, elementary charge, Coulomb constant and speed of light respectively.\\ \indent
        3.  Unit cell parameters of NPP crystal: a, b, c, $\beta$ and its other parameters that have given in section
        2.\\ \indent
        4.  L: crystal thickness that in our simulation it is 3$\mu$m.\\
        And the outputs of
        $MATLAB$ program are: $n_{x}$, $n_{y}$ in each wavelength and their related errors from experimental data.\\ \indent
        System calibration is done in this method that we obtain two refractive indexes in two wavelengths
        with $\epsilon$ (eccentricity), u (semimajor axis of ellipse) and Z (equivalent positive charge) in a
        way that refractive indexes in two wavelengths are very close to experimental data.
        Then we would see that refractive index in other wavelengths
        with same $\epsilon$, u and Z will be achieved. Of course these values, $\epsilon$, u and
        Z would be close to experimental structure of crystal, for example u would be greater than and smaller than
        minimum and maximum sizes of six lengths of benzene hexagonal respectively, or $\epsilon$ would be small
        and greater than zero. In other hand these values must be logical. From this method in our simulation we have
        obtained $\epsilon=0.26, Z=3.9, u=1.4A^{0}$ that is very close to experimental and structural data .
        Tables (\ref{Table2}) and (\ref{Table3})
        \begin{table*}
             \centering
             \caption{Comparison of experimental data and simulation results for $n_{x}$}\label{Table2}
             \begin{tabular}{|c|c|c|c|c|c|c|c|c|c|c|}
             \hline
             $Wavelength(\lambda-nm)$     & 509 & 532 & 546& 577 & 589 & 633 & 644 & 690 & 1064 & 1340 \\ \hline\hline
             $n_{x}(Experimental)\cite{3}$ & 2.355 & 2.277 & 2.231 & 2.153 & 2.128 & 2.066 & 2.055 & 2.051 & 1.926 & 1.917\\ \hline
             $n_{x}(Simulation)$ & 2.290 & 2.271 & 2.284 & 2.195 & 2.200 & 2.176 & 2.150 & 2.113 & 1.910 & 1.806\\ \hline
             $Error\%$ & 2.7 & 0.23 & 2.4 & 1.9 & 3.3 & 5.3 & 4.6 & 3 & 0.8 & 5.8\\ \hline
             \end{tabular}
        \end{table*}
        \begin{table*}
             \centering
             \caption{Comparison of experimental data and simulation results for $n_{y}$}\label{Table3}
             \begin{tabular}{|c|c|c|c|c|c|c|c|c|c|c|}
             \hline
             $Wavelength(\lambda-nm)$     & 509 & 532 & 546& 577 & 589 & 633 & 644 & 690 & 1064 & 1340 \\ \hline\hline
             $n_{y}(Experimental)\cite{3}$ & 2.116 & 2.024 & 1.982 & 1.927 & 1.911 & 1.876 & 1.857 & 1.857 & 1.774 & 1.757\\ \hline
             $n_{y}(Simulation)$ & 2.068 & 2.052 & 2.033 & 2.019 & 2.004 & 1.974 & 1.944 & 1.913 & 1.733 & 1.656\\ \hline
             $Error\%$ & 2.3 & 1.4 & 2.6 & 4.8 & 4.9 & 5.2 & 4.7 & 3 & 2.3 & 5.7\\ \hline
             \end{tabular}
        \end{table*}
        compare the results of simulation with experimental data. From these tables we perceive $n_{x}$ and $n_{y}$ that
        we attained from simulation, have $5.8$ percent maximum error and 0.23 percent minimum error.
        These values are acceptable measures. Figures \ref{Fig12} and \ref{Fig13} compare these data with Sellmeier
        data that are provided from \cite{3,5}.
        \section{Conclusion}
        The suggested physical model in this paper that is based on assumption of energetic particles could be a
        powerful tool for analyzing and explaining processes that happen in waveguides in microscopic sizes.
        From similar models we can explain optical events such as second harmonic generation and optical
        polarization in fiber optic applications and optical communication. We will try to continue applying
        this method for justification of another phenomenon. Our approach is quasi-quantum and its advantages
        are:\\ \indent
        1-We may to obtain scientific data of a real material,\\ \indent
        2-We utilize momentary quantities in time domain,\\ \indent
        3- Our method may predict some events in nanostructurs and very short time by considering short-range
        interatomic forces and its effect on photons.
        \bibliographystyle{IEEEtran}
        \bibliography{pa85per8asl25}

 \end{document}